\documentclass[aps,pre,superscriptaddress,showpacs,twocolumn,floatfix]{revtex4}

\usepackage{graphicx}

\def\u0{\underline{0}}

\begin{document}
\title{Minimizing  energy below the glass thresholds}

\author{Demian Battaglia}
\email[]{battagli@sissa.it}
\affiliation{SISSA, Via Beirut 9, I-34100 Trieste, Italy,}

\author{Michal Kol\'{a}\v{r}}
\email[]{kolarmi@sissa.it}
\affiliation{SISSA, Via Beirut 9, I-34100 Trieste, Italy,}

\author{Riccardo Zecchina} 
\email[]{zecchina@ictp.trieste.it}
\affiliation{ICTP, Strada Costiera 11, I-34100 Trieste, Italy.}

\begin{abstract}
Focusing on the optimization version of the random K-satisfiability
problem, the MAX-K-SAT problem, we study the performance of
the finite energy version of the Survey Propagation (SP) algorithm.
We show that a simple (linear time) backtrack decimation strategy is
sufficient to reach configurations well below the lower bound
for the dynamic threshold energy and very close to the analytic
prediction for the optimal ground states.  A comparative numerical
study on one of the most efficient local search procedures is also
given.

\end{abstract}
\maketitle

\section{Introduction}
The problem of finding variable configurations
that minimize the energy of a system with competitive interactions has
been and still is a central one in the study of
complex systems, like spin glasses in physics, protein folding and
regulatory networks in biology, and optimization problems in computer
science (see \textit{e.g.,} \cite{review_dynamics,Proteins,Clustering,
optiphysics, newoptiphysics}).

Among the tools for numerical investigations of complex systems at low
temperatures the simulated annealing (SA) algorithm \cite{SA} and its
variants have played a major role. Such stochastic processes satisfy
detailed balance and their behavior can be compared with static and
dynamical mean-field calculations. However, in problems in which the
interest is focused on zero temperature ground states and where the
proliferation of metastable states causes an exponential slowdown in the
equilibration rate, the applicability of SA-like algorithms is limited to
relatively small system sizes.

In computer science the field of combinatorial optimization
\cite{Papadimitriou} deals precisely with the general issue of classifying the
computational difficulty (``hardness'') of minimization problems and
of designing search algorithms. Similarly to statistical physics models, 
a generic combinatorial optimization problem is
composed of many discrete variables---\textit{e.g.,} Boolean variables, finite
sets of colors or Ising spins---which interact through constraints
typically involving a small number of variables, that in turn sum up
to give the global cost-energy function.

When the problem instances are extracted at random from nontrivial
ensembles (that is ensembles which contains many instances that are
hard to solve), computer science meets physics in a very direct way:
many of the models considered to be of basic interest for Computer
Science are nothing but spin glasses defined over finite connectivity
random graphs, the well studied diluted spin glasses \cite{TCS,AI}. Their 
associated energy function counts the number
of violated constraints in the original combinatorial problem (with
ground states corresponding to optimal solutions).  Understanding the
onset of hardness of such systems is at the same time central to
computer science and to $T=0$ statistical physics with surprisingly
concrete engineering applications.  For instance, among the most
effective error correcting codes and data compression methods are the
Low Density Parity Check algorithms
\cite{Spielman,RichardsonUrbankeIntroduction,Sourlas}, which indeed
implement an energy minimization of a spin glass energy defined over a
sparse random graph.  In such problems, the choice of the graph
ensemble is a part of the designing techniques, a fact that makes spin
glass theory directly applicable.

The above example is however far from representing the general
scenario for combinatorial problems: in many situations the
probabilistic set up is not defined and, consequently, the notion of
typical-case analysis does not play any obvious role.  The study of
the connection (if any) between worst-case and typical-case complexity
is indeed an open one and very few general results are known
\cite{Ajtai}.  Still, a precise understanding of non-trivial random
problem instances promises to be important under many aspects.  New
algorithmic results as well as many mathematical issues have been put
forward by the statistical physics studies, with examples ranging from
phase transitions \cite{Nature,Science} and out-of-equilibrium
analysis of randomized algorithms \cite{Physical_Methods} to new
classes of message-passing algorithms \cite{MZ,BMZ}.

The physical scenario for the diluted spin glasses version of hard
combinatorial problems predicts a trapping in metastable states for
exponentially long times of local search dynamic process satisfying
detailed balance.  Depending on the models and on the details of the
process---\textit{e.g.,} cooling rate for SA---the long time dynamics is
dominated by different types of metastable states at different
temperatures \cite{Montanari:Ricci:1}.  A common feature is that at zero
temperature and for simulation times which are sub-exponential in the
size of the problem there exists an extensive gap in energy which
separates the blocking states from true ground states.

Such behavior can be tested on concrete random instances which
therefore constitute a computational benchmark for more general
algorithms. Of particular interest for computer science are randomized
search processes which do not properly satisfy detailed balance and that are
known (numerically) to be more efficient than SA-like algorithms in
the search for ground states \cite{Randomized_Algorithms}.  Whether
the physical blocking scenario applies also to these artificial
processes, which are not necessarily characterized by a proper Boltzmann distribution at
long times, is a difficult open problem. The available numerical
results and some approximate analytical calculations
\cite{Semerjian_Monasson,Barthel_et_al} seem to support the existence of a thermodynamical gap, a
fact which is of up-most importance for optimization.  For this reason
(and independently from physics), during the last decade the problem
of finding minimal energy configurations of random combinatorial
problems similar to diluted spin-glasses---\textit{e.g.,} random
K-Satisfiability (K-SAT) or Graph Coloring---has become a very
popular algorithmic benchmark in computer science \cite{AI}.

In the last few years there has been a great progress in the study of
spin glasses over random graphs which has shed new light on mean-field
theory and has produced new algorithmic tools for the study of low
energy states in large single problem instances.  Quite surprisingly,
problems which were considered to be algorithmically hard for local
search algorithms, like for instance random K-SAT close to a phase
boundary, turned out to be efficiently solved by the Survey
Propagation (SP) algorithm arising from the replica symmetry broken
(RSB) cavity approach to diluted spin glasses.  Such type of results
calls for a rigorous theory of the functioning of SP (which is a non
local process) and bring new mathematical challenges of potential
practical impact.

Scope of this paper is to display a set of new numerical and
algorithmic results which complete previously published results on the
SP algorithm. We shall deal only with the random K-SAT problem even
though we expect the algorithmic outcomes to be applicable to other
similar problems like, for instance, the random graph coloring.

The paper is organized as follows.  In Sections \ref{KSAT}, \ref{SP}
we briefly review the known results on random K-SAT together with the
SP equations over single instances at finite pseudo-temperature. We
discuss as well in \ref{SP-Y} how the SP algorithm can be modified in
order to study the region of parameters with finite ground state
energy (UNSAT phase), where not all constraints of the
underlying random K-SAT problem can be satisfied simultaneously.  In
Sec. \ref{results} we discuss then the performance of SP as an
optimization device. At variance with the SAT phase in which many
clusters of zero energy configurations coexist and where SP works
efficiently without need of correcting variable assignments, in the
UNSAT phase an efficient implementation of SP requires the
introduction of---at least---a very simple form of backtracking
procedure (similar to the one proposed in
\cite{Parisi_backtrack}).  We show that a linear time backtrack
is enough to reach energies compatible with those predicted by the
analytic calculations in the infinite size limit in the relevant
region of parameters.  We give moreover numerical evidence for the
existence of threshold states for one of the most efficient randomized
local search algorithms for solving random K-SAT, namely WalkSat
\cite{WalkSat}. We display a blocking mechanism at an energy level
which is definitely above the lower bound for the dynamical threshold
states predicted by the stability analysis of the 1-RSB cavity
equations.  Finally, for the deep UNSAT phase, we report on numerical
data on convergence times for both WalkSat and SA which are in
agreement with the predicted existence of full RSB (f-RSB) phases.
Conclusions and perspectives are briefly discussed in Sec.~\ref{Conclusions}.

\section{Brief review of random K-SAT}\label{KSAT}
K-SAT is a NP-complete problem \cite{Garey_Johnson} (for $K>2$) which
lies at the root of combinatorial optimization. It is very easy to
state: Given $N$ Boolean variables and $M$ constraints taking the form
of clauses, {\it K-SAT consists in asking whether it exists an
assignment of the variables that satisfies all constraints}.  Each
clause contains exactly $K$ variables, either directed or negated, and
its truth value is given by the OR function.  Since the same variable
may appear directed or negated in different clauses, competitive
interactions among clauses may set in.

As mentioned in the introduction, in the last decade there has been a
lot of interest on the random version of K-SAT: for each
clause the variables are chosen uniformly at random (with no
repetitions) and negated with probability $1/2$.

In the large $N$ limit, random K-SAT displays a very interesting
threshold phenomenon.  Taking as control parameter the ratio of
number of clauses to number of variables, $\alpha=M/N$, there exists 
a phase transition at
a finite value  $\alpha_c(K)$ of this ratio.  For $\alpha<\alpha_c(K)$ the
generic problem is satisfiable (SAT), for $\alpha>\alpha_c(K)$ the
generic problem is not satisfiable (UNSAT).

This phase transition has been seen numerically
\cite{kirkpatrick:selman:94} and it is of special
interest since extensive experiments \cite{AI} have shown
that the instances which are algorithmically hard to solve are exactly
those where $\alpha$ is close to $\alpha_c$.  Therefore, the study of
the SAT/UNSAT phase transition is considered of crucial
relevance for understanding the onset of computational complexity in
typical instances \cite{TCS}.  A lot of work has been focused on the study of
both the decision problem (\textit{i.e.,} determining with a YES/NO answer
whether a satisfying assignment exists), and the
optimization version in which one is interested in minimizing the
number of violated clauses when the problem is UNSAT (random MAX-K-SAT problem).

On the analytical side, there exists a proof that the threshold
phenomenon exists at large $N$ \cite{Friedgut}, although the fact that
the corresponding $\alpha_c$ has a limit when $N \to \infty$ has not
yet been established rigorously.  Upper bounds
$\alpha_{\mathrm{UB}}(K)$ on $\alpha_c$ have been found using first
moment methods \cite{dubois_kirousis} and variational interpolation
methods \cite{guerra_franz:leone:03}, and lower bounds
$\alpha_{\mathrm{LB}}(K)$ have been found using either explicit
analysis of some algorithms \cite{lowbound1}, or some second moment
methods \cite{achlioptas:moore:02}. For random MAX-K-SAT theoretical bounds are also known
\cite{Achlioptas_Naor_Peres,max-3-sat-7/8}, as well as rigorous results on the
running times of random walk and approximation algorithms \cite{Schoning,Ben-Sasson,Parkes}.

Recently, the cavity method of statistical physics has been applied to K-SAT
 \cite{Science,MZ,MMZ} and the thresholds have been
computed with high accuracy.  A lot of work is going on in order to
provide a rigorous foundation to the cavity results and we refer to
\cite{MMZ} for a more complete discussion of these aspects.

In what follows we shall concentrate on the $K=3$ case and we will be
interested in analyzing the behavior of different algorithms in the
region of parameter in which the random formulas are expected to be
hard to solve or to minimize.  The energy function which is used in
the zero temperature statistical mechanics studies is taken proportional to
the number of violated clauses in a given problem so that a zero
energy ground state corresponds to a satisfying assignment.  The
energy of a single clause is positive (equals 2 for technical reasons)
if the clause is violated and zero if it is satisfied. The overall
energy is obtained by summing over clauses and reads
\begin{equation}\label{hamiltonian}
E=2\sum_a\frac{\prod_{i=1}^{3}\left(1+J_{a,i} s_i^a\right)}{2}
\label{energy}
\end{equation}
where $s^a_i$ is the $i$-th binary (spin) variable appearing in clause
$a$ and the coupling $J_{a, i}$ takes the value 1 (resp. -1) if the
corresponding variable appears not negated (resp. negated) in clause
$a$.  For instance the clause $(x_1 \vee \bar x_2 \vee x_3)$
has an energy $\frac{1}{4} (1+s_1) (1-s_2) (1+s_3)$ where the Boolean
variables $x_i=\{0,1\}$ are connected to the spin variables by the
transformation $s_i=(-1)^{x_i}$.

The phase diagram of the random 3-SAT problem as arising from the
statistical physics studies can be very briefly summarized as
follows. 

For $\alpha < 3.86$, the $T=0$ phase is at zero energy (the problem is
SAT). The entropy density is finite and the phase is Replica Symmetric
(RS) and unfrozen. Roughly speaking, this means that there exists one
giant cluster of nearby solutions and that the effective fields vanish
linearly with the temperature.  

For $3.86 < \alpha < 3.92$, there is a full RSB phase. The solution
space breaks in clusters and the order parameter becomes a nested
probability measure in the space of probability distribution
describing cluster to cluster fluctuations.  The phase is still SAT
and unfrozen \cite{Biroli_et_al,Parisi_SAT04}. 

At $\alpha \simeq 3.92$ there is a discontinuous transition toward a
clustered frozen phase \cite{Science,MZ}.  Up to $\alpha =4.15$ the
phase is f-RSB while above the 1-RSB solution becomes
stable\cite{Montanari:Parisi:Ricci}.  The {\it complexity},
that is the normalized logarithm of the number of clusters, is finite
in this region.  At finite energy there exist even more metastable
states which act as dynamical traps. The 1-RSB metastable states
become unstable at some energy density $E_G(\alpha)$ which constitutes
a lower bound to the true dynamical \textit{threshold energy} (see
Sec.~\ref{SP} for more details).  

At $\alpha=4.2667$ the ground state energy becomes positive and
therefore the typical random 3-SAT problem becomes UNSAT. At the same
point the complexity vanishes. The phase remains 1-RSB up to
$\alpha=4.39$ where an instability toward a zero complexity full RSB
phase appears. 

In the region $4.15 < \alpha < 4.39$, the 1-RSB ansatz for the ground state is stable
against higher orders of RSB, but the 1-RSB predictions become
unstable for energies larger than the \textit{Gardner energy}. The
instability line intersects with the 1-RSB
ground state extimation at
the two extremes of the interval, inside which it provides a lower
bound to the true threshold energy (see
Ref.~\cite{Montanari:Parisi:Ricci} for a
comprehensive discussion).

Further (preliminary) f-RSB corrections suggest that the true threshold
states have energies very close to the lower bound and hence the
interval $A=[4.15,4.39]$ should be taken as the region where to take
really hard benchmarks for algorithm testing.
\begin{figure}
\includegraphics[width=\columnwidth]{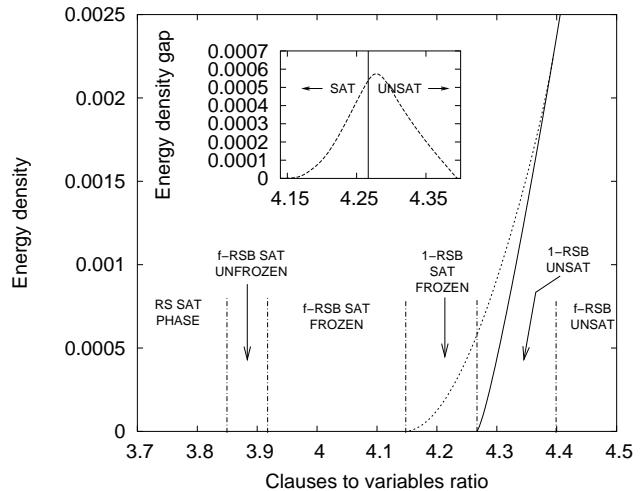}
\caption{ The solid line is an extimation for the ground state energy,
while the dashed curve represents the Gardner energy, providing a
lower bound for the threshold states (numerical data adapted from
ref. \cite{Montanari:Parisi:Ricci}).  In the inset we show that the
difference between the Gardner and the ground state energy is strictly
positive in the small 1-RSB stable region around the SAT/UNSAT
transition critical point (indicated by the vertical line): it is
expected that it is hard for heuristics based on local search to find
assignments inside the closed area delimited by the energy gap curve.}
\label{small_gap_fig}
\end{figure}
As displayed in Fig.~\ref{small_gap_fig}, the actual value of the
energy gap is very small close to the end points of $A$. In order to
avoid systematic finite size errors, numerical simulations should be
done close to the SAT/UNSAT point, \textit{i.e.,} far from the end point of $A$.
Consistently with the fact that finite size fluctuations are
relatively big ($O(\sqrt{N}))$, even close to $\alpha_c$ problem
sizes of the order at least of $N=10^5$ are necessary in order to
observe a matching with the analytic predictions.

\section{Brief review of SP equations}\label{SP}
The 1-RSB cavity equations which have been used to study the typical
phase diagram of random K-SAT become the SP equations once
reformulated to run over single problem instance \cite{MZ}. This is
done by avoiding the averaging process with respect to the underlying
random graphs.  Thanks to the self-averaging property of the random
K-SAT free energy \cite{Self-averaging}, the SP equations can be used
both to re-derive the phase diagram of the problem and, more
important, to access detailed information of algorithmic relevance
about a given problem instance.  In particular, the SP equations
provide information about the statistical behavior of the single
variables in the stable and metastable states of given energy density.

The 1-RSB cavity equations are iterative equations (averaged over the
disorder) for the probability distribution functions (pdf) of
effective fields that describe their cluster-to-cluster fluctuations.
The order parameter is a probability measure in the space of pdf's; it
tells the probability that a randomly chosen variable has a certain
associated pdf in states at a given energy density.

In SP and more in general in the cavity approach, one assumes to know
pdf's of the fields of all variables in the temporary absence of one
of them. Then one writes the induced pdf of the local field acting on
this ``cavity'' variable in absence of some other variable interacting
with it (\textit{i.e.,} the so called Bethe lattice approximation for
the problem).  These relations define a closed set of equations for
the pdf's that can be solved iteratively.  The equations are exact if
the cavity variables acting as inputs are uncorrelated, \textit{e.g.,}
over trees, or are conjectured to be an asymptotically exact
approximation over locally tree-like structures \cite{MZ} where the
typical distance between randomly chosen variables diverges in the
large $N$ limit (as $\ln N$ for diluted random graphs).  The full list
of the cavity fields over the entire underlying graph, in the SP
implementation, constitutes the order parameter.  From the cavity
fields one may determine the total field acting on each variable in
all metastable states of given energy density and this information can
be used for algorithmic purposes.

\begin{figure}
\includegraphics[width=0.62\columnwidth]{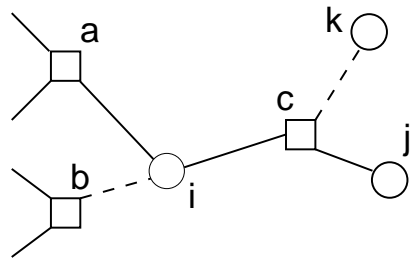}
\caption{Factor graph representation. Variables are represented by
  circles, and are connected by function nodes, represented by
  squares; if a variable appears negated in a clause, the connecting
  line is dashed.}
\label{factorgraph}
\end{figure}

A clear formalism for the single sample analysis is given by the
factor graph representation \cite{factor_graph} of K-SAT: variables
are represented by $N$ circular ``variable nodes'' labeled with
letters $i,j,k,\ldots$ whereas the K-body interactions are represented by
$M$ square ``function nodes'' (carrying the clause energies) labeled
by $a,b,c,\ldots$ (see Fig.~\ref{factorgraph})

For random 3-SAT, function nodes have connectivity $3$, variable nodes
have a Poisson connectivity of average $3 \alpha$ and the overall
graph is bipartite.  The total energy is nothing but the sum of
energies of all function nodes as given by Eq. (\ref{energy}).
 
Adopting the message-passing notation and strictly following
\cite{MZ}, we call $u$-messages the contribution to the cavity
fields coming from the different connected branches of the graph.  In
SP the messages along the links of the factor graph have a functional
nature carrying information about distributions of $u$-messages over
the states at a given value of the energy, fixed by a Lagrange
multiplier $y$: we call these distributions of messages
$u$-surveys. The SP equations can be written at any ``temperature''
(the inverse of the Lagrange multiplier $y$ is actually a
pseudo-temperature, see \cite{MZ}). However they acquire a particularly
simple form in the limit $1/y\to 0$, which is the limit of interest
for optimization purposes, at least in the SAT region.

In K-SAT, the $u$-surveys are parameterized by two real numbers and SP
can be implemented very efficiently.  Each edge $a \to i$, from a
function node $a$ to a variable node $i$, carries a $u$-survey $Q_{a
\to i}(u)$. From these $u$-surveys one can compute the cavity fields
$h_{i\to b}$ for every $b\in V(i)$, which in turn determine new output
$u$-surveys (see Fig.~\ref{popdynfig}).

Very schematically, the SP equations can be implemented as follows.
Let $V(i)$ be the set of function nodes connected to the variable $i$,
$V(a)$ the set of variables connected to the function node $a$; let
us denote by $V(i)\setminus a$ and $V(a)\setminus i$ the same sets
deprived respectively of the clause $a$ and of the variable $i$. Given
then a random initialization of all the $u$-surveys $Q_{a \to i}(u)$,
the function nodes are selected sequentially at random and the
$u$-surveys are updated according to a complete set of coupled
functional equations (see Fig.~\ref{popdynfig} for the notation):
\begin{figure}
\includegraphics[width=0.9\columnwidth]{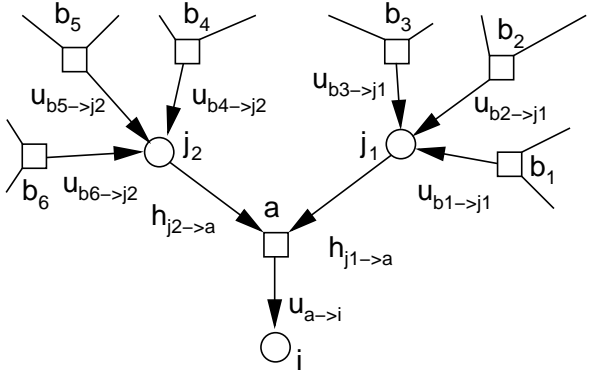}
\caption{Cavity fields and $u$-messages. The $u$-survey for the
    $u$-message $u_{a\to i}$ depends on the pdf's of the cavity fields $h_{j_1\to
    a}$ and $h_{j_2\to a}$. These are on the other side dependent on
    the $u$-surveys for the $u$-messages incoming to the variables
    $j_1$ and $j_2$.}\label{popdynfig}
\end{figure}
\begin{eqnarray}
P_{j \to a}(h_{j\to a})&=&C_{j \to a} \int \mathcal{D}Q_{j,a}\,\,\delta
  \Big(h-\sum_{b\in V(j)\setminus a} u_{b\to j} \Big) \nonumber\\ 
\times \exp \Big(y
  &\!\!\!\big(\vert&\!\!\!\!\!\!\!\!\sum_{b \in V(j)\setminus a} u_{b\to j} \vert - \sum_{b\in
  V(j)\setminus a} \vert u_{b\to j} \vert\big) \Big),
\label{P}\\
Q_{a \to i}(u)&=&\int \mathcal{D}P_{a,i}\,\,
\delta\left( u-\hat{u}_{a\to i}\left(\left\{h_{j\to a}\right\}\right)\right),
\label{Q2}
\end{eqnarray}
where the $C_{i \to a}$'s are normalization constants, the
function $\hat{u}_{a\to i}$ is:
\begin{equation}
\hat{u}_{a\to i}\left(\left\{h_{j\to
a}\right\}\right)=J_{a,i}\prod_{j\in V(a)\setminus
i}\theta\left(J_{a,j}h_{j\to a}\right),
\end{equation}
and the integration measures are given by:
\begin{equation}
\mathcal{D}Q_{j,a} = \prod_{b\in V(j)\setminus a}Q_{b \to j}
(u_{b\to j})\,du_{b\to j},
\end{equation}
\begin{equation}
\mathcal{D}P_{a,i} = \prod_{j\in V(a)\setminus i}P_{j\to
    a}(h_{j\to a})\,dh_{j\to a}.
\end{equation}

Parameterizing the $u$-surveys as
\begin{equation}
Q_{a \to i}(u)=\eta_{a\to i}^0\delta(u)+\eta_{a
\to i}^+\delta(u-1)+\eta_{a \to i}^-\delta(u+1)
\end{equation}
where $\eta_{a\to i}^0=1-\eta_{a\to i}^+ - \eta_{a\to i}^-$, the above
set of equations (\ref{P},\ref{Q2}) defines a non-linear map over the
$\eta$'s.

Once a fixed point is reached, from the list of the $u$-surveys one may
compute the  normalized pdf  of the \textit{local field} acting on each variable:
\begin{eqnarray}
P_{i}(H)&=&C_i \int \mathcal{D}\widehat{Q}_i\,\,\delta
\Big( H -\sum_{b \in V(i)} u_{b\to i} \Big)\nonumber\\
\times&\!\!\exp&\!\!\Big(y\big(\vert \!\!\! \sum_{b
\in V(i)} u_{b\to i} \vert - \sum_{b \in V(i)} \vert u_{b\to i} \vert\big) \Big),
\label{local_field}\\
\mathcal{D}\widehat{Q}_i &=& \prod_{b\in V(i)}Q_{b \to i}
(u_{b\to i})\,du_{b\to i}.
\end{eqnarray}
It should be remarked that
$P_i(H)$ is in general different from the family of \textit{cavity
  fields} pdf's $P_{i\to b}(h)$ computed by mean of (\ref{P}).

From the knowledge of the cavity and local fields pdf's, one
derives the (Bethe) free energy at the level of 1-RSB:
\begin{equation}
\Phi(y)= \frac{1}{N}\left(\sum_{a=1}^M \Phi^f_{a}(y)-\sum_{i=1}^N
\Phi^v_i(y) (\Gamma_i-1)\right) \ ,
\label{freeonesamp1}
\end{equation}
where $\Gamma_i$ is the connectivity of the variable $i$ and:
\begin{widetext}
\begin{eqnarray}
\Phi^f_{a}(y)&=&-\frac{1}{y} 
\ln \left\{
\int \prod_{i \in V(a)}\mathcal{D}Q_{i,a}\,\,
\exp \left[ -y \min_{\{ \sigma_i,i \in V(a)\} } \left( E_a -\sum_{i
\in V(a)} \left[ \sum_{b\in V(i)\setminus a}u_{b\to i} \right] \sigma_i +
\sum_{b\in V(i)\setminus a} \vert u_{b\to i} \vert \right) \right] \right\}
,\nonumber \\
\Phi^v_i(y)&=&-\frac{1}{y} \ln \left \{ 
\int \mathcal{D}\widehat{Q}_i\,\, \exp\left[y (\vert
\sum_{a \in V(i)} u_{a\to i} \vert- \sum_{a \in V(i)}\vert u_{a\to i} \vert )
\right] \right \}=-\frac{1}{y} \ln (C_i).
\label{freeonesamp2}
\end{eqnarray}
\end{widetext}
Here, $E_a$ is the energy contribution of the function node $a$. The
maximum value of the free-energy functional provides a lower bound
estimation of the ground state energy of the Hamiltonian
(\ref{hamiltonian}) defined on the sample. In the SAT region the
free-energy functional $\Phi(y)$ is always non positive and it is
increasing in the limit $y\to\infty$; in the UNSAT region, on the
contrary, it exhibits a positive maximum for $y=y^*$ (see \cite{MZ}).

From the free-energy density of a given instance, it is
straightforward to compute numerically its complexity
$\Sigma(y)=\partial \Phi(y) / \partial(1/y)$ and its energy density
$\epsilon(y)=\partial(y\Phi(y))/\partial y$.  We remind that the
complexity is linked to the number of pure states (\textit{i.e.,}
clusters of configurations) of energy $E$, by the defining relation
$\mathcal{N}(E)=\exp\left(N\Sigma(E)\right)$. The energy level
represented by the largest number of configurations, $e_{th}$, is given by:
\begin{equation}
\Sigma(e_{th})=\max_{E}\left(\Sigma(E)\right).
\end{equation}
Further RSB corrections may be needed to locate the precise value of
$e_{th}$, which is in any case lower bounded the largest energy of
1-RSB stable states, the so called {\it Gardner energy} $E_G$.  It is
expected that local search strategies get trapped at energies close,
but not necessarily equal, to the threshold energy (see
refs. \cite{Montanari:Ricci:1} for a throrough discussion on the role
of the iso-complexity states~\cite{iso}).  More elaborated strategies
not properly satisfying detailed balance (\textit{e.g.,} WalkSat for
the K-SAT problem) could in principle overcome this type of barriers;
however, the available numerical and analytical results suggest that
also these more sophisticated randomized searches undergo an
exponential slowdown, with different layers of states acting as
dynamical traps, depending on the details of the heuristics.

\section{SP in the UNSAT region}\label{SP-Y}
In the SAT phase, where the $y \to \infty$ limit is taken, the
convolutions (\ref{P}) filter out completely any clause-violating
truth value assignment.  This feature is extremely useful for
satisfiable formulas, but it becomes undesired when our sample is
presumably unsatisfiable. 

In the UNSAT region the SP equations require a finite value of the
Lagrange multiplier $y$. The
filtering action of the exponential re-weighting term in (\ref{P}) is
then weakened and the messages computed by the SP equations can 
vehicle information pointing to states with a non vanishing number of
violated constraints.

\subsection{The finite pseudo-temperature recursive equations} 
The SP equations simplify considerably in the $y\to \infty$ limit and
lead to extremely efficient algorithmic implementations, as discussed
in great detail in \cite{BMZ}.  In the case of finite pseudo-temperature 
$1/y$ the same simplification cannot take place because of
the presence of a nontrivial re-weighting factor. Still, a relatively
fast recursive procedure can be written.  Let us consider a variable
$j$ having $\Gamma_j$ neighboring function nodes and let us compute the
cavity field pdf $P_{j \to a}(h)$ where $a\in V(j)$. We start by
randomly picking up one function node in $V(j)\setminus a$, denoted as
$b_1$, and we calculate the following ``$h$-survey'':
\begin{equation} \label{pdrecinit}
P_{j\to a}^{(1)}(h)= \eta_{b_1\to i}^{0}\,\delta(h) +
\eta_{b_1\to i}^{\/+}\,\delta(h-1) + \eta_{b_1\to i}^{\/-}\,\delta(h+1).
\end{equation}
The function $P_{j\to a}^{(1)}(h)$ would correspond to the true local
field pdf of the variable $j$ in the case in which $b_1$ was the only
neighboring clause (as denoted by the upper index).

The following steps of the recursive procedure consist
in adding the contributions of all the other function nodes in
$V(j)\setminus a$, clause by clause (Fig.~\ref{popdynrecfig}):
\begin{figure}
\includegraphics[width=0.9\columnwidth]{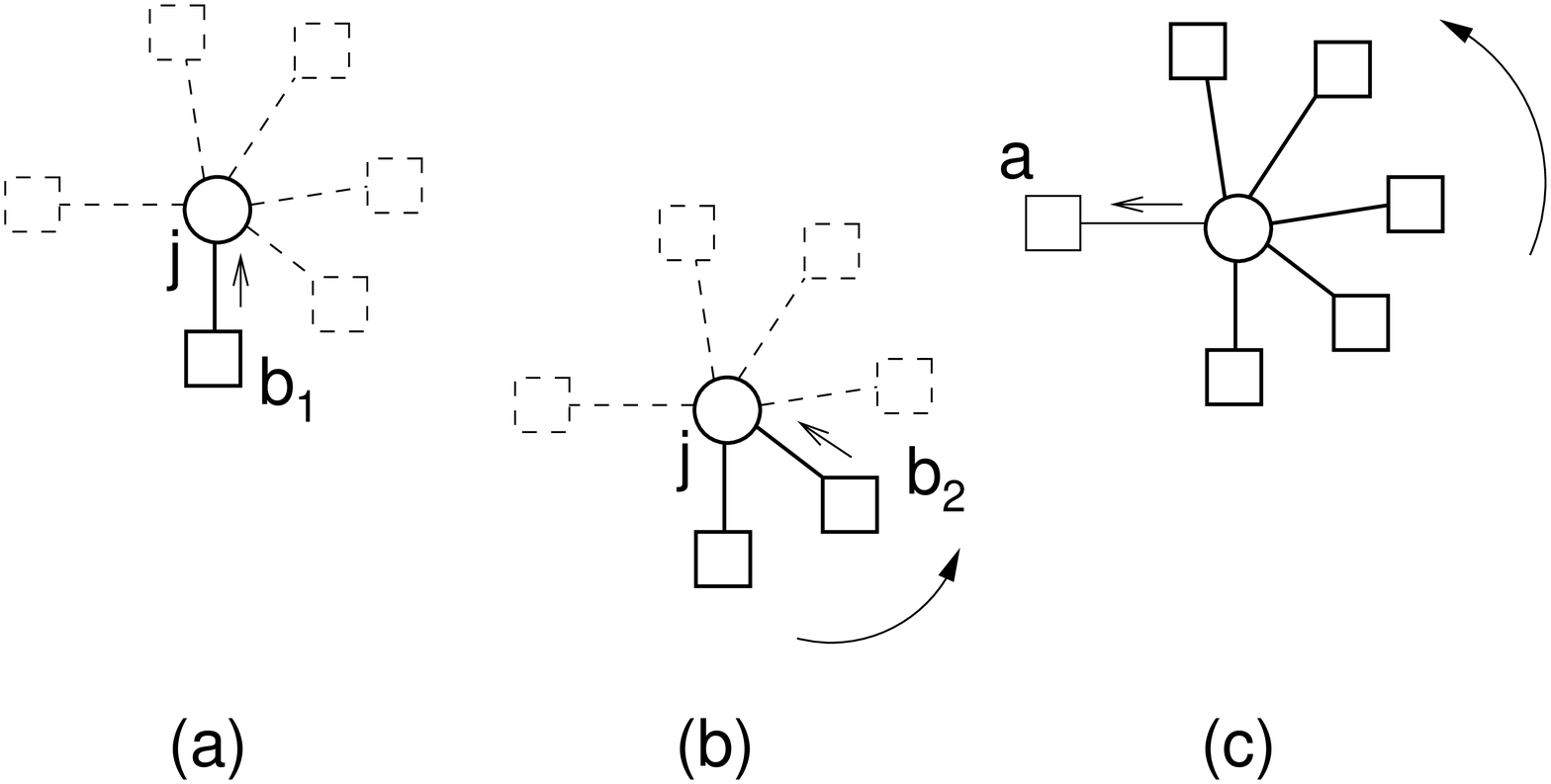}
\caption{Computing recursively a cavity pdf. (a) In order to find a
  single cavity pdf $P_{j\to a}(h)$, a single clause $b_1$ in
  $V(j)\setminus a$ is picked up at random and the $u$-survey
  $Q_{b_1\to j}$ is used to compute equation (\ref{pdrecinit}); (b)
  The contributions of all the other function nodes in $V(j)\setminus
  a$ are then added, clause by clause; (c) The pdf computed
  recursively after $\Gamma_j -1$ iterations coincides with $P_{j\to
  a}(h)$.}\label{popdynrecfig}
\end{figure}
\begin{eqnarray} \label{pdrec}
  \widetilde{P}_{j\to a}^{(\gamma)}(h)&=&
  \eta_{b_\gamma\to j}^{0}\,\widetilde{P}_{j\to a}^{(\gamma-1)}(h)\\
  &+& \eta_{b_\gamma\to j}^{\/+}\,\widetilde{P}_{j\to
      a}^{(\gamma-1)}(h-1)\,\exp\left [-2y\,\hat{\theta}(-h)\right]\nonumber\\ 
  &+& \eta_{b_\gamma\to j}^{\/-}\,\widetilde{P}_{j\to
      a}^{(\gamma-1)}(h+1)\,\exp\left [-2y\,\hat{\theta}(h)\right].\nonumber
\end{eqnarray}
Here $\widetilde{P}^{(\gamma)}_{j\to a}(h)$ is an unnormalized pdf and
$\hat{\theta}(h)$ is a step function equal to $1$ for $h \geq 0$ and
zero otherwise.  The recursion ends after $\gamma = \Gamma_j-1$ steps,
when the influence of every clause in $V(j)\setminus a$ has been taken
in account. The final cavity-field pdf $P_{j\to a}(h)$ can be found
straightforwardly by computing the pdf $\widetilde{P}_{j\to
a}^{(\Gamma_j-1)}(h)$ for all values of the field $-\Gamma_j + 1 < h <
\Gamma_j -1$ and by normalizing it.

As already pointed out in Section~\ref{SP}, the knowledge of $K - 1$
input cavity-field pdf's can be used to obtain a single output
$u$-survey.  Let us compute for instance the $u$-survey $Q_{a\to
i}(u)$ (see always Fig.~\ref{popdynfig} for the notation).  In order
to do that, we need first the cavity field pdf's $P_{j\to a}(h)$ for
every $j\in V(a)\setminus i$. The parameters $\{\eta_{a\to i}^0,
\eta_{a\to i}^+, \eta_{a\to i}^-\}$ are then updated according to the
formulas:
  \begin{equation}\label{equpdateeta}
    \eta_{a\to i}^{J_{a, i}} = \prod_{n=1}^{K-1} W_{j_n\to
      a}^{J_{j_n, a}}, \quad \eta_{a\to i}^{-J_{a, i}} = 0, \quad
    \eta_{a\to i}^0 = 1 - \eta_{a\to i}^{J_{a, i}},
  \end{equation}
where we introduced the weight factors:
\begin{equation}\label{Wplus}
  W_{j\to a}^+ = \sum_{h = 1}^{\Gamma_j - 1} P_{j\to a}(h),\quad  W_{j\to a}^- = \sum_{h = -\Gamma_j + 1}^{-1} P_{j\to a}(h).
\end{equation}
It should be remarked that $Q_{a\to i}(u)$ depends only on one single
nontrivial $\eta_{a\to i}^{J_{a, i}}$ (from now simply referred to as
$\eta_{a\to i}$). We could say that a single kind of message can be
produced, telling the receiver literal to assume the truth value
``TRUE''; this message is transmitted along the edge $a\to i$ with a
probability $\eta_{a\to i}$, corresponding to the probability that the
only way of not violating the constraint $a$ is to set appropriately
the truth value of $i$.

Starting from a full collection of $u$-surveys at a given time, it is
possible to realize a complete update of all the parameters
$\left\{\eta_{a\to i}\right\}$ by systematical application of the
recursions (\ref{pdrecinit}), (\ref{pdrec}) and of the relation
(\ref{equpdateeta}); from the new set of $u$-surveys, new cavity field
pdf's can be computed and the procedure continues until when
self-consistence of $\eta$'s is reached. This procedure can be
efficiently implemented numerically and allows us to determine the
fixed point of the population-dynamics equations (\ref{P}),
(\ref{Q2}), for a general value of $y$.

\subsection{The SP-Y algorithm} 
In the usual SP-inspired decimation \cite{BMZ}, the computation
of the local field pdf's $P_{i}(H)$ is used to decide a truth value
assignment for the most biased variables.  Indeed, it is reasonable
that a spin tends to align itself with the most probable direction of
the local field.  A ranking can be realized by finding all the
probability weights 
\begin{equation}\label{Wplus_loc}
 W_{j}^+ = \sum_{H = 1}^{\Gamma_j} P_{j}(H),\quad   W_{j}^- = \sum_{H = -\Gamma_j}^{-1} P_{j}(H),
\end{equation}
and by sorting the variables according to the values of a bias function:
\begin{equation}\label{bias_SP}
b_{\mathrm{fix}}(j) =  |W_{j}^+ - W^-_j|.
\end{equation}
The local field pdf's $P_{j}(H)$ can be naturally calculated resorting
to the iterations (\ref{pdrecinit}), (\ref{pdrec}): computation is
done simply by sweeping over the whole set of neighboring function
nodes $V(j)$, including also the contribution of the skipped edge
$a\to j$.  By fixing in the right direction the spin of the most
biased variable, we actually reduce the original $N$ variable problem
to a new one with $N-1$ variables.  New $u$-surveys are then
computed. Doing that we have to take care of fixed variables: if $i$
is fixed, its cavity field pdf's must be of the form:
\begin{equation}\label{fully_polarized}
P_{i\to a}(h) = \delta\left(h-J_{a, i}s_{i}\right);
\end{equation}
regardless of the recursions (\ref{pdrecinit}), (\ref{pdrec}). The
complete polarization reflects the knowledge of the truth value of the
literals depending on the spin $s_i$.

The procedure of decimation continues until when a full truth
assignment has been generated or until when convergence has been lost
or a paramagnetic state has been reached; in the latter cases the
original formula is simplified according to the partial truth
assignment already generated and the simplified formula is passed to a
specialized heuristic. Our choice of preference is the WalkSat
algorithm \cite{WalkSat}, which is by far more efficient than SA in
the hard region of the 3-SAT problem, as we have checked exhaustively.
Very briefly, the strategy of WalkSat is the following one: at each
time step the current assignment is changed by randomly alternating
greedy moves (where the variable which maximizes the number of
satisfied clauses if fixed) and  random-walk steps (in which a
variable belonging to a randomly chosen unsatisfied clause is selected
and flipped). WalkSat stops if either a satisfying assignment is found
or if the maximum number of allowed spin flips (the ``cutoff'') is
reached (see Ref. \cite{Orponen} for another recently analyzed and very
efficient heuristics).

When working at finite pseudo-temperature, we have to take in account
the possibility that some non optimal fixing is done in presence of
thermal ``noise''.  After several updates of the $u$-surveys some
biases of fixed spins may become smaller than the value they had at
the time when the corresponding spin was fixed. Certain local fields can
even revert their orientation.  Small or positive values of an index
function like:
\begin{equation}\label{bias_SP-bk}
b_{\mathrm{backtrack}}(j) =  -s_j\left(W_{j}^+ - W^-_j\right),
\end{equation}
can track the appearance of such dangerous fixed spins and this
information can be used to implement some ``error removal'' procedure;
for instance, a simple strategy can be devised where both unfixing and
fixing moves are performed at a fixed ratio $0 \leq r < 0.5$ (see
\cite{Parisi_backtrack} for another backtracking implementation).

The actual SP with finite $y$ simplification procedure (SP-Y) 
will depend not only on the
backtracking fraction $r$, but even more on the choice of the inverse
pseudo-temperature $y$. The simplest possibility is to keep it fixed
during the simplification, but one may choose to dynamically update
it, in order to stay as close as possible to the maximum $y^*$ of the
free energy functional $\Phi(y)$ (which corresponds to select the
ground state in the 1-RSB framework, as we have seen in Section~\ref{SP}).

\begin{figure}
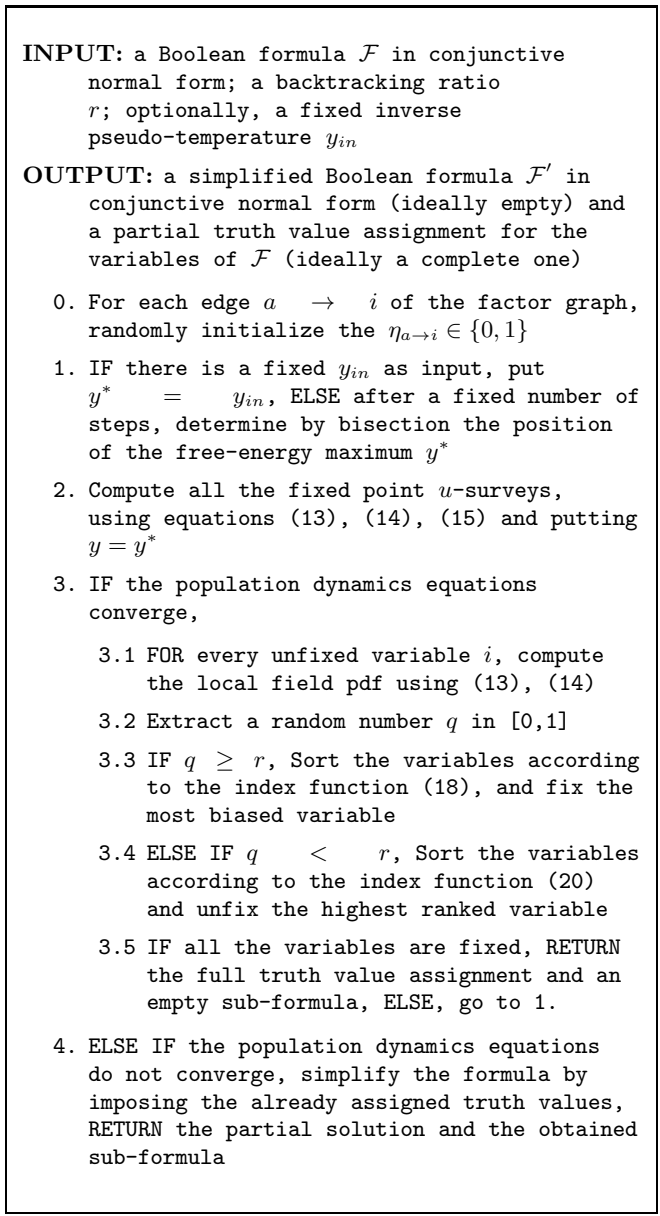

\framebox{
\begin{minipage}{0.95\columnwidth}
\vspace{1em}
{\tt
\begin{description}
\item[INPUT:] a Boolean formula $\mathcal{F}$ in conjunctive normal
form; a backtracking ratio $r$; optionally, a fixed inverse
pseudo-temperature $y_{in}$

\item[OUTPUT:] a simplified Boolean formula $\mathcal{F}'$ in
conjunctive normal form (ideally empty) and a partial truth value
assignment for the variables of $\mathcal{F}$ (ideally a complete one)
\end{description}
\begin{itemize}
\item[0.] For each edge $a \to i$ of the factor graph,
  randomly initialize the $\eta_{a \to i} \in \{ 0,1\}$
\item[1.] IF there is a fixed $y_{in}$ as input, put $y^* = y_{in}$, ELSE after a fixed number of steps, 
determine by bisection the position of the free-energy maximum $y^*$
\item[2.] Compute all the fixed point $u$-surveys, using equations
  (\ref{pdrecinit}), (\ref{pdrec}), (\ref{equpdateeta}) and putting $y = y^*$
\item[3.] IF the population dynamics equations converge, 
        \begin{itemize}
        \item[3.1] FOR every unfixed variable $i$, compute the local
        field pdf using (\ref{pdrecinit}), (\ref{pdrec})
        \item[3.2] Extract a random number $q$ in [0,1]
        \item[3.3] IF $q \geq r$, Sort the variables according to the index function
          (\ref{bias_SP}), and fix the most biased variable
        \item[3.4] ELSE IF $q < r$, Sort the variables according to
        the index
        function (\ref{bias_SP-bk}) and unfix the highest ranked variable       
        \item[3.5] IF all the variables are fixed, RETURN the full
        truth value assignment and an empty sub-formula, ELSE, go to 1. 
        \end{itemize}
\item[4.] ELSE IF the population dynamics equations do not converge,
  simplify the formula by imposing the already assigned truth values,
  RETURN the partial solution and the obtained sub-formula
\end{itemize}
}
\vspace{1em}
\end{minipage}
}
\caption{The SP-Y simplification algorithm.}\label{pseudofig}
\end{figure}

The
equations (\ref{freeonesamp1}), (\ref{freeonesamp2}) can be rewritten
in the following form, suitable for numerical computation:
\begin{eqnarray}\label{Phi_SP-Y_clause}
\Phi_{a}^f(y) = -\frac{1}{y}\Big[&\!\!\!\ln&\!\!\!\Big(1+\left(\mathrm{e}^{-y}-1\right)
\prod_{i\in V(a)}W_{i\to a}^{J_{a, i}}\Big)\nonumber\\
&-&\!\!\ln\Big(\prod_{i\in V(a)}C_{i\to a}\Big)\Big],\\
\label{Phi_SP-Y_var}
\Phi_{i}^v(y) = -\frac{1}{y}&\!\!\!\ln&\!\!\!\left(C_i\right).
\end{eqnarray}
In Fig.~\ref{pseudofig} we give a summary of the simplification
procedure in a standard pseudo-code notation.
The first release of the SP-Y code can be downloaded from
\cite{SP-Y}.

\section{Optimizing the energy below the threshold states}\label{results}
As we have already discussed in Section~\ref{SP}, it is expected that, in the
thermodynamical limit, any local search algorithm gets trapped in the
vicinity of exponentially numerous threshold states with energy
$e_{th}$ and that any local heuristics is in general unable to find
the optimal assignment in the thermodynamical limit. To verify this
prediction, we conducted various experiments, both in the SAT and in
the UNSAT phase, focusing on the comparison between the WalkSat
heuristics performance after and before different kinds of SP-Y
simplification. In most of the situations, we decided to analyze
carefully single large-sized samples instead of a larger number of
smaller problems: we verified in fact that the sample-to-sample
fluctuations tend to be unrelevant for size of order $10^4$ and
larger.

\subsection{SAT region}
\begin{figure}
\center
\includegraphics[angle=-90, width=1.00\columnwidth]{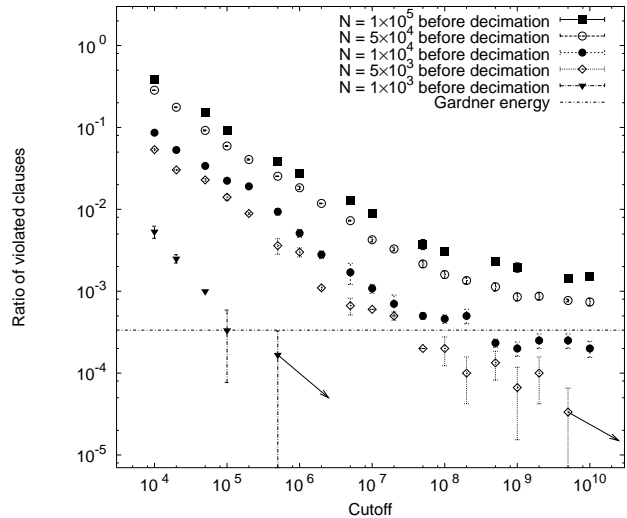}
\caption{Threshold energy effect in SAT region. The WalkSat
performance for various samples of different sizes and $\alpha = 4.24$
is presented. With increasing size, the curves appear to saturate
above the Gardner energy. An arrow indicates that the next data point
corresponds to a SAT assignment.}\label{Fig01}
\end{figure}    
The aim of the first set of experiments was to check the actual
existence of the threshold effect.  We ran WalkSat over different
formulas in the hard-SAT region, with fixed $\alpha = 4.24$ and sizes
varying between $N=10^3$ and $N=10^5$, reaching a maximum cutoff of
$10^{10}$ spin flips. The obtained results are plotted in
Fig.~\ref{Fig01}; the Gardner energy is also reported for comparison
with the data.  Even if for small-size samples the local search
algorithm is able to find a SAT assignment, for larger formulas ($N
\sim \mathcal{O}(10^4)$) WalkSat does not succeed in reaching the
ground state, its relaxation profile suffers of critical slowdown, and
saturates at some well defined level. This is actually expected,
because the Gardner energy becomes $\mathcal{O}(1)$ only for
$N\sim10^4$ or larger, and for a smaller number of variables the
threshold effect should be negligible when compared to finite size
effects.

We remind that WalkSat cannot be considered as an equilibrium
stochastic process and that it is not possible to infer that its
saturation level coincides with the sample threshold energy; we can
anyway claim that WalkSat is unable to explore the full energy
landscape of the problem, and that the enormous number of non optimal
valleys is unavoidably hiding the true ground states. Plateaus in the
relaxation profiles of WalkSat have indeed been already discussed in
\cite{Semerjian_Monasson,Barthel_et_al} and ascribed to metastable
states acting as dynamical traps.

For the $N=10^4$ formula a trapping effect becomes clearly visible in
our experiments, but the saturation plateau is below the Gardner lower
bound. The finite-size fluctuations are still of the same order of the
energy gap between the ground and the threshold states and the
experimental conditions are distant from the thermodynamical
limit. When the size is increased up to $10^5$ variables, the
saturation level moves finally between the full RSB lower bound and
the 1-RSB upper bound for $e_{th}$.

\begin{figure}
\center
\includegraphics[angle=-90, width=1.00\columnwidth]{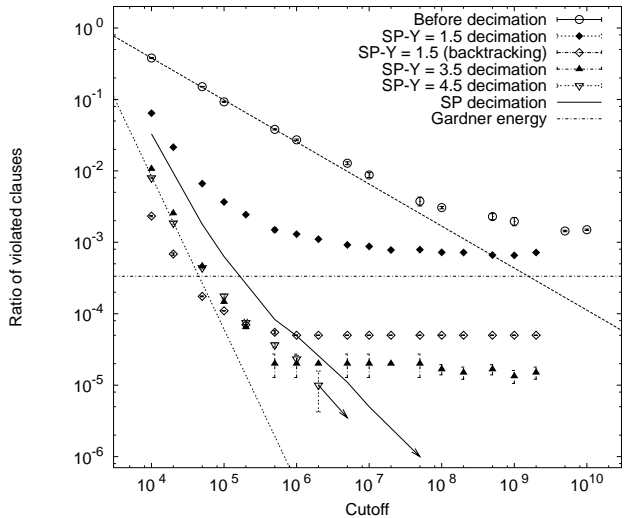}
\caption{Efficiency of SP-Y in the SAT region (single sample with
$N=10^5$ variables and $\alpha = 4.24$). After SP-Y simplification,
WalkSat is generally able to find solutions below the Gardner
threshold; in some cases, it succeeds even in finding complete
satisfying assignment. An arrow indicates that the next data point
corresponds to a SAT assignment.}\label{Fig02}
\end{figure} 

The efficiency of the SP-Y simplification strategy against the glass
threshold is discussed in Fig.~\ref{Fig02}.  We simplified a single
randomly generated formula ($N=10^5$, $\alpha=4.24$) at several fixed
values of pseudo-temperature.  The solid line shows for comparison the
WalkSat results after a standard SP decimation (\textit{i.e.,} $y\to\infty$):
the ground state, $E = 0$, is reached as expected, after a rather
small number of spin flips.  The same happens after SP-Y
simplifications performed at a large enough inverse pseudo-temperature
($y>4$); one should remind indeed that in the SAT region the optimal
value for $y$ would be infinite, and that in that limit the SP-Y
recursions reduce to the SP equations.  After simplification with
smaller $y$'s, the WalkSat cooling curves reach again a saturation
level, which is nevertheless below the Gardner energy, unless $y$ is
too small: the threshold states of the original formula have not been
able to trap the local search, even if the ground state becomes
inaccessible.  As we have indeed already discussed, working at finite
temperature increases the probability of violating a clause when doing a
spin fixing, and this is particularly evident in the SAT region where
every assignment that does not satisfy some constraint should be
filtered out.

The procedure is intrinsically error prone, and it will allow in
general to reach only ``good states'', but not the true optimal
solutions (the smaller the parameter $y$, the higher the saturation
level will be).  As we shall discuss in the next section, the use of
backtracking partially cures the accumulation of errors at finite y:
the saturation level can in fact be significantly lowered by keeping
the same pseudo-temperature and introducing a small fraction of
backtrack moves during the simplification. In Fig.~\ref{Fig02} the 
data for $y=1.5$ shows the importance of backtracking. While the run of SP-Y
without backtracking has led to a plateau above Gardner energy,
with the introduction of backtrack moves we find energies well below the threshold.

\subsection{UNSAT region}
When entering the UNSAT region, the task of looking for the optimal
state becomes harder. The expected presence of violated constraints in
the optimal assignments really forces us to run the simplification at
a finite pseudo-temperature. Unfortunately, after many spin fixings,
the recursions (\ref{pdrecinit}), (\ref{pdrec}) stop to converge for
some finite value of $y$ before the maximum of the free energy is
reached, most likely because the sub-problem has entered a full RSB
phase.  At this point one should switch to a 2-RSB version of SP which
we did not realize, yet.  Alternatively, one could try to run directly
the final heuristic search (hoping that the full RSB sub-system is not
exponentially hard to optimize) or more simply one may continue the
decimation process by selecting the largest $y$ for which the
computation converge. We decided to implement the latter choice until
either convergence is lost independently from the value of $y$ or a
paramagnetic state is reached.

\begin{figure}
\center
\includegraphics[angle=-90, width=1.00\columnwidth]{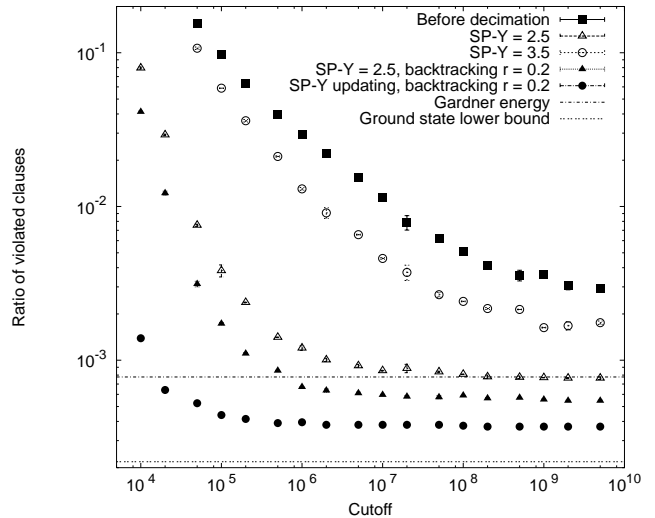}
\caption{SP-Y performance in the UNSAT region (single sample with
$N=10^5$ variables and $\alpha = 4.29$).  Several simplification
strategies are compared; the need for backtracking is readily visible,
and its introduction allows to reach energies closer to the ground
state than to the Gardner lower bound.}\label{Fig03}
\end{figure}

In our experiments we studied several 3-SAT sample problems belonging
to the 1-RSB stable UNSAT phase. We employed WalkSat as an example of
standard well-performing heuristics. Although WalkSat is not optimized
for unsatisfiable problems, in the 1-RSB stable UNSAT region 
it performs still much better than any basic implementation of SA. 
We observed anyway that, even after
$10^{10}$ spin flips, the WalkSat best assignments were still quite
distant from the Gardner energy, for various samples of different size
and $\alpha$. In Fig.~\ref{Fig03} we show the results relative to many
different SP-Y simplifications with various values of $y$ and $r$ for a
single sample with $N=10^5$ and $\alpha=4.29$.
\begin{figure}
\center
\includegraphics[angle=-90, width=1.00\columnwidth]{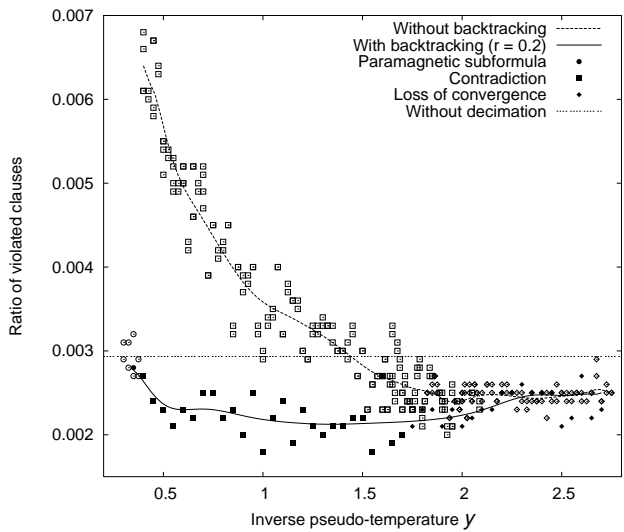}
\caption{Backtracking efficiency. Many SP-Y simplifications of a single
sample with $N = 10^4$ variables and $\alpha = 4.35$ have been
performed at fixed but different values of pseudo-temperature; the
introduction of a small fraction of backtracking moves eliminates
essentially the need for a time consuming optimization of the
parameter $y$. The empty points refer to simplifications without
backtracking, the full points to simplifications with a backtracking
ratio $r=0.2$. A diamond indicates that the simplification process
stopped because of loss of convergence, a circle because of finding a
completely unbiased paramagnetic state, and the squares indicates that
the loss of convergence happened at an advanced stage where some
clause-violating assignments have already been introduced by
SP-Y.}\label{Fig04}
\end{figure}
The simplification produced always an improvement in the WalkSat
performance, but, in absence of backtracking, we were unable to go
below the Gardner lower bound (although we touched it in some cases:
in Fig.~\ref{Fig03} we show the data for a simplification at fixed
$y=2.5$; a simplification with runtime optimization of $y$ reached the
same level).

The relative inefficiency of these first attempts of simplification
was not due to the threshold effect alone, but also to an extreme
sensitivity to the choice of $y$, as pointed out by a second set of
experiments making use of backtracking.  We performed first an
extensive analysis of the simultaneous optimization of $y$ and $r$,
using smaller samples in order to produce more experimental
points. After some trials, the fraction $r=0.2$ appeared to be the
optimal one, at least for our implementation, and in the small region
under investigation of the K-SAT phase diagram. The data in
Fig.~\ref{Fig04} refers to a formula with $N=10^4$ variables and
$\alpha=4.35$. The dashed horizontal line shows the WalkSat best
energy  obtained on the original formula after $10^9$ spin flips. The
WalkSat performance was seriously degraded when simplifying at too
small values of $y$, but the introduction of backtracking cured the
problem, identifying and repairing most of the wrong assignments. The
WalkSat efficiency became actually almost independent from the choice
of pseudo-temperature, whereas in absence of error correction a time
consuming parameter tuning was required for optimization.

Coming back to the analysis of the sample of Fig.~\ref{Fig03},
the backtracking simplifications allowed us to access states
definitely below the Gardner lower bound. The combination of runtime
$y$-optimization and of error correction was even more effective: after
a rather small number of spin flips, WalkSat reached a saturation
level strikingly closer to the ground state lower bound than to the
Gardner energy.  A further valuable effect of introduction of the
backtracking was the increased efficiency of the formula simplification
itself: in the backtracking experiments, SP-Y was able to determine a
truth value for more than 80\% of the variables before losing
convergence, while without backtracking, the algorithm stopped on
average after only 40\% of fixings.

All the samples analyzed in the previous sections were taken from the
1-RSB stable region of the 3-SAT problem, where the equations
(\ref{P}), (\ref{Q2}) are considered to be exact.  For $\alpha >
4.39$, the phase becomes full~RSB and SP loses convergence before the
free energy $\Phi(y)$ reaches its maximum from the very first step of
the decimation procedure.  While a full~RSB version of SP would most
likely provide very good results, SP-Y still can be used in a
sub-optimal way by selecting the largest value of $y$ for which
convergence is reached.  Numerical experiment show that indeed the
performance of SP-Y are in good agreement with the analytical
expectations.  However, it should be noticed that in this region the
use of SP is not necessary.  Although the performance of WalkSat and
SA can be improved by the SP simplification, the SA alone is already
able of finding close-to-optimum assignments efficiently (as expected
for a full~RSB scenario) and behaves definitely better than WalkSat.

\section{Conclusions}\label{Conclusions}
In this paper, we have displayed the performance of SP as an
optimization device and shown that configurations well below the
threshold states can be found efficiently.  Similar results are
expected to hold also for random satisfiable instances very close to
the critical point for which the combined use of finite
pseudo-temperature and backtracking could give access to the SAT
optima.

It would be of some interest to analyze further improvements of the
decimation strategies as well as to consider more structured factor
graphs within a variational framework, in which some correlations can
be put under control.

A possible application of SP-Y--like algorithms can be found in
information theory: lossy data compression based on Low Density
Parity Check schemes leads to optimization problems which are 
indeed very similar to the one discussed in this paper.

\section{Acknowledgments}
We thank A. Braunstein, M. M\'ezard, G. Parisi and F. Ricci-Tersenghi
for very fruitful discussions.
This work has been supported in part by
the European Community's Human Potential Programme under contract
HPRN-CT-2002-00319, STIPCO.

\end{document}